\documentclass[journal]{IEEEtran}

\usepackage{graphicx,cite,epsfig,amssymb,amsmath,subfigure,url,stfloats,latexsym}
\usepackage{array}
\usepackage{arydshln}
\usepackage{amsfonts}
\usepackage{pgfplots}
\usepackage{algorithm}
\usepackage[noend]{algpseudocode}
\usepackage{epstopdf}
\usepackage{amsfonts,amsthm}
\usepackage{multirow}
\usepackage{mathrsfs}
\usepackage{subfigure}
\usepackage{geometry}

\definecolor{mygray}{gray}{.9}
\graphicspath{{figures/}}
\newcolumntype{C}[1]{>{\PreserveBackslash\centering}p{#1}}
\newcolumntype{R}[1]{>{\PreserveBackslash\raggedleft}p{#1}}
\newcolumntype{L}[1]{>{\PreserveBackslash\raggedright}p{#1}}

\usepackage[flushleft]{threeparttable}

\begin{document}

\title{Finite Field Multiple Access for Sourced Massive Random Access with Finite Blocklength}

\author{{Qi-yue Yu,~\IEEEmembership{Senior Member,~IEEE}, Shi-wen Lin, and Shu Lin,~\IEEEmembership{Life Fellow,~IEEE}}
\thanks{The work presented in this paper was supported by the National Natural Science Foundation of China under Grand No. 62071148.}
}

\maketitle

\begin{abstract}
For binary source transmission, this paper introduces the concept of element-pair (EP) and establishes that when the Cartesian product of $J$ distinct EPs satisfies the unique sum-pattern mapping (USPM) structural property, these $J$ EPs can form a uniquely-decodable EP (UD-EP) code. EPs are treated as virtual resources allocated to different users in finite fields, serving to distinguish users. This approach enables the reordering of multiplexing and channel encoding modules, effectively addressing the finite blocklength (FBL) challenge in multiuser reliable transmission.
Next, we introduce an orthogonal EP code $\Psi_{\rm o, B}$ constructed over an extension field GF($2^m$). Using this EP code, we develop a time-division mode of finite-field multiple-access (FFMA) systems, consisting of sparse-form and diagonal-form structures.
Based on the diagonal-form (DF) structure, we present a specific configuration, referred to as polarization-adjusted DF-FFMA, which can simultaneously obtain the power gain and coding gain from the entire blocklength.
Simulation results demonstrate that the proposed FFMA systems significantly improve error performance over a Gaussian multiple-access channel, compared to a slotted ALOHA system.
\end{abstract}


\vspace{-0.1in}
\section{Introduction}
For the next generation of wireless communication, ultra massive machine type communication (um-MTC) is required to serve a huge number of users (or devices) simultaneously, 
e.g., the connection density of an um-MTC system is around $10^{6} \sim 10^{8}$ devices/km$^2$ \cite{6G, UMA_PZ, UMA_2022}. 
The required multiple-access (MA) technique should simultaneously support massive users with short packet traffic, and achieve an acceptable {per-user probability of error (PUPE)} \cite{MIT_2017}.

In general, grant-free random access (RA) is considered as a promising MA scheme for supporting um-MTC communications \cite{SourcedRA_1,Yu_UDAS}. The grant-free random access technique can be classified into sourced and unsourced RA, which can be applied in different scenarios \cite{SourcedRA_2}.
In this paper, we focus on the sourced RA, which is used to serve device-oriented scenarios, e.g., monitoring the status of sensors, enabling the base station (BS) to be aware of both the messages and users (or devices) identities \cite{SourcedRA_1}.
For a sourced RA system, each user is assigned a unique signature to identify the active user(s) at the BS. Typically, we can use different resources, sequences and/or codebooks to identity different users.
The classical sourced RA system utilizes time slots to distinguish users, 
operating in a {slotted ALOHA} scheme, which simplifies both the active user detection and multiuser detection algorithms \cite{ALOHA_1}.

Nevertheless, for the massive random access scenario, the reliable performance of the slotted ALOHA scheme is limited by the blocklength of each time slot \cite{Capacity_GMC_2017, Capacity_GMC_2021}.
The blocklength or the number of degrees of freedom (DoF) is set to be $N$, which is also called \textit{finite blocklength (FBL)} constraint.
Suppose there are totally $J$ users. To separate users without ambiguity, the number of time slots is set to be equal to the number of users $J$. Thus, the blocklength of each time slot is equal to $N/J$. 
According to \cite{Capacity_GMC_2021}, we know that the maximal rate of each time slot is equal to
\begin{equation*}
C(P) - \sqrt{\frac{V(P)}{(N/J)}} Q^{-1}(\epsilon) + \frac{1}{2} \frac{\log (N/J)}{(N/J)} + {\mathcal O}(\frac{1}{(N/J)})
\end{equation*}
where $C(P)$ is the channel capacity, $P$ is the maximum power constraint, and $\epsilon$ is the average error probability.
$V(P) = \frac{P(P+2)}{2(1+P)^2}$ is the dispersion of the Gaussian channel.
Obviously, the maximal rate of each time slot decreases with the increased number of users $J$, especially $J$ is a large value.
Hence, it is challenging to design suitable schemes for solving the FBL of massive users reliability transmission problem.

In this paper, an MA coding is devised based on the \textit{element-pair (EP) code} constructed over finite fields (FFs) \cite{FFMA}. 
We refer the proposed MA technique based on EP-coding as \textit{finite-field MA (FFMA)} technique (or coding). An MA communication system using FFMA-coding is referred to as an FFMA system for simplicity. 
Unlike the slotted ALOHA scheme, which operates with a blocklength of $N/J$, the proposed FFMA system leverages the entire blocklength $N$ to achieve error correction capabilities across all DoF.
The key distinction lies in the order of encoding and multiplexing (MUX) modules in the proposed FFMA system, which differs from conventional MA techniques. 
It is noted that the finite-field MUX is performed by the EP-coding.
By performing multiplexing before channel coding, the system maximizes channel gains from all available DoF.



The rest of this paper is organized as follows. Section II introduces the concept of EP-coding and presents an orthogonal EP code constructed over GF($2^m$). 
The encoding of the orthogonal EP code is introduced in Section III. Section IV presents an FFMA system for a massive MA transmission. 
In Section V, simulations of the error performances of the proposed FFMA systems are given. Section VI concludes the paper with some remarks.


\vspace{-0.1in}

\section{Orthogonal EP Code over GF($2^m$)}

Suppose GF($q$) is a finite-field with $q$ elements, where $q$ is a prime number or a power of the prime number. Let $\alpha$ denote a primitive element of GF($q$), and the powers of $\alpha$, i.e.,
$\alpha^{-\infty} = 0, \alpha^0=1, \alpha, \alpha^2, \ldots, \alpha^{(q - 2)}$, 
give all the $q$ elements of GF($q$). 
For binary source transmission, a transmit bit is either $(0)_2$ or $(1)_2$.
We utilize two different elements in a finite-field GF($q$) to express the bit information, 
i.e., $(0)_2 \mapsto \alpha^{l_{j,0}}$ and $(1)_2 \mapsto \alpha^{l_{j,1}}$, 
where $\alpha^{l_{j,0}}, \alpha^{l_{j,1}} \in$ GF($q$) and $l_{j,0} \neq l_{j,1}$. 
We define the selected two elements $(\alpha^{l_{j,0}}, \alpha^{l_{j,1}})$ as an \textit{element pair (EP)}.

In this paper, we construct \textit{orthogonal element pair (EP) codes} over GF($2^m$), 
where $q = 2^m$ and $m$ is a positive integer with $m \ge 2$.
Then, each element $\alpha^{l_{j}}$, with $l_{j} = 0, 1, \ldots, 2^m - 2$, in GF($2^m$) can be expressed as a linear sum of $\alpha^0 = 1, \alpha, \alpha^2, \ldots, \alpha^{(m - 1)}$ with coefficients from GF($2$) as 
\begin{equation} \label{e2.2}
    \alpha^{l_{j}} = a_{j,0} + a_{j,1} \alpha + a_{j,2} \alpha^2 + \ldots + 
    a_{j,m-1} \alpha^{(m-1)}.      
\end{equation}
From (\ref{e2.2}), we see that the element $\alpha^{l_{j}}$ can be uniquely represented by the $m$-tuple $(a_{j,0}, a_{j,1}, \ldots, a_{j,m-1})$ over GF($2$), which is a linear combination of 
$\alpha^0, \alpha^1, \ldots, \alpha^{m-1}$, i.e., $\alpha^{l_{j}} = \oplus_{i=0}^{m-1} a_{j,i} \alpha^i$.
For $0 \le i < m$, it is known that $\alpha^i = (0, 0, \ldots, 1, 0,\ldots, 0)$ is an $m$-tuple with a 1-component at the $i$-th location and $0$s elsewhere. 

Based on the finite-field GF($2^m$), we define $m$ EPs as
\begin{equation*}
  \begin{array}{ll}
    C_1 = (0, \alpha^{0})  &= \alpha^{0} \cdot C_{\rm B} =\{{\bf 0}, (1, 0, \ldots, 0)\}, \\
    C_2 = (0, \alpha^{1})  &= \alpha^{1} \cdot C_{\rm B} =\{{\bf 0}, (0, 1, \ldots, 0)\}, \\
    \quad \vdots \\
    C_{m} = (0, \alpha^{m-1}) &= \alpha^{m-1} \cdot C_{\rm B} =\{{\bf 0}, (0, 0, \ldots, 1)\}, 
  \end{array}
\end{equation*}
where ${\bf 0}$ is an $m$-tuple of zeros, and $C_{\rm B} =(0,1)$.
The subscript ``B'' stands for ``binary''.
Hence, for $1 \le j \le m$, the $j$-th EP $C_j$ is given as $C_j = \alpha^{j-1} \cdot C_{\rm B}$.
The $m$ EPs together can form an EP set, defined by ${\Psi}_{\rm o,B}$, i.e., 
\begin{equation} \label{Psi_oB}
  \begin{aligned}
  {\Psi}_{\rm o,B} &= \{C_1, C_2, \ldots, C_m\}, \\
                   &= \{\alpha^{0} \cdot C_{\rm B}, \alpha^{1} \cdot C_{\rm B}, \ldots, 
                   \alpha^{m-1} \cdot C_{\rm B}\}.
  \end{aligned}
\end{equation}  
With $0 \le i < m$, any two different EPs in ${\Psi}_{\rm o,B}$, e.g., $C_j$ and $C_{j'}$ where $j \neq j'$, are \textit{orthogonal} to each other.

Let $(u_1, u_2, \ldots, u_j, \ldots, u_J)$ be a $J$-tuple over GF($2^m$) in which the $j$-th component $u_j$ is an element from the EP $C_j$, where $1 \le j \le J$ and $J \le m$. 
The $J$-tuple $(u_1, u_2, \ldots, u_J)$ is an element in the \textit{Cartesian product} $C_1 \times C_2 \times \ldots \times C_J$ of the EPs in the EP set $\Psi_{\rm o,B}$. 
We view each $J$-tuple ${\bf u} = (u_1, u_2, \ldots, u_j, \ldots, u_J)$ in
$C_1 \times C_2 \times \ldots \times C_J$ as a $J$-user \textit{EP codeword}.
In our paper, ${\Psi}_{\rm o,B}$ stands for an EP set and also an EP code.
Hence, the Cartesian product
\begin{equation*} 
\Psi_{\rm o,B} \triangleq (\alpha^0 \cdot C_{\rm B}) \times (\alpha^1 \cdot C_{\rm B}) \times \ldots \times 
(\alpha^{m-1} \cdot C_{\rm B}),
\end{equation*}
of the $m$ EPs in $\Psi_{\rm o,B}$ can form an $m$-user \textit{EP code} over GF($2^m$) with $2^m$ codewords.

Let $(u_1, u_2, \ldots, u_{J})$ and $(u_1', u_2', \ldots, u_{J}')$ be \textit{any two} $J$-tuples in $C_1  \times C_2 \times \ldots \times C_{J}$ of the EP code ${\Psi}_{\rm o,B}$. 
If $\oplus_{j=1}^{J} u_j \neq \oplus_{j=1}^{J} u_j'$, 
then an \textit{finite-field sum-pattern (FFSP)} $w$ uniquely specifies a $J$-tuple in $C_1 \times C_2 \times \ldots \times C_{J}$. 
That is to say the mapping 
\vspace{-0.1in}
\begin{equation} \label{e.UDmap}
(u_1, u_2, \ldots, u_{J}) \Longleftrightarrow w = \bigoplus_{j=1}^{J} u_j,
\end{equation} 
is a one-to-one mapping. 
In this case, given the FFSP $w = \bigoplus_{j=1}^{J} u_j$, 
we can uniquely recover the $J$-tuple $(u_1, u_2, \ldots, u_{J})$ \textit{without ambiguity}. 
We say that the Cartesian product $C_1 \times C_2 \times \ldots \times C_{J}$ has a \textit{unique sum-pattern mapping (USPM) structural property}, and form an \textit{uniquely-decodable EP (UD-EP)} code.

Because of the orthogonality between any two EPs in $\Psi_{\rm o,B}$, it is easy to know the EP codeword in $\Psi_{\rm o,B}$ has USPM structural property.
We call ${\Psi}_{\rm o,B}$ a $J$-user \textit{orthogonal uniquely decodable EP (UD-EP) code} over GF($2^m$). 
When this code is used for an MA communication system with $J$ users, the $j$-th component $u_j$ in a codeword $(u_1, u_2, \ldots, u_{J})$ is the element to be transmitted by the $j$-th user.

\vspace{-0.1in}
\section{Encoding of Orthogonal UD-EP Codes}

In this section, we first introduce the encoder of an EP code $\Psi_{\rm o,B}$.
Then, we investigate the orthogonal UD-EP codes $\Psi_{\rm o,B}$ encoded by a channel encoder ${\mathcal C}_{gc}$. The subscript ``gc'' of ${\mathcal C}_{gc}$ stands for ``globe code'', since the channel code ${\mathcal C}_{gc}$ is used through the MA transmission.

\vspace{-0.1in}
\subsection{Binary to Finite-field GF($q$) Transform Function}

Let the bit-sequence at the output of the $j$-th user be ${\bf b}_j = (b_{j,0}, b_{j,1},\ldots, b_{j,k}, \ldots, b_{j,K-1})$, where $K$ is a positive integer, $1 \le j \le J$ and $0 \le k < K$.
The EP encoder is to map each bit-sequence ${\bf b}_j$ uniquely into an element-sequence 
${\bf u}_j = (u_{j,0},u_{j,1},\ldots, u_{j,k},\ldots, u_{j,K-1})$ 
by a \textit{binary to finite-field GF($q$) transform function} ${\rm F}_{{\rm B}2q}$, 
i.e., $u_{j,k} = {\rm F}_{{\rm B}2q}(b_{j,k})$.

Assume each user is assigned an EP, e.g., the EP of $C_j = (\alpha^{l_{j,0}}, \alpha^{l_{j,1}})$ is assigned to the $j$-th user for $1 \le j \le J$.
The subscript ``$j$'' of ``$l_{j,0}$'' and ``$l_{j,1}$'' stands for the $j$-th EP,
and the subscripts ``$0$'' and ``$1$'' of ``$l_{j,0}$'' and ``$l_{j,1}$'' represent the input bits are $(0)_2$ and $(1)_2$, respectively. 
For $1 \le j \le J$, we can set the $k$-th component $u_{j,k}$ of ${\bf u}_j$ as
\begin{equation} \label{F_b2q}
  u_{j,k} = {\mathrm F}_{{\mathrm B}2q}(b_{j,k}) \triangleq b_{j,k} \odot C_j = 
  \left\{
    \begin{aligned}
      \alpha^{l_{j,0}}, \quad b_{j,k} = 0  \\
      \alpha^{l_{j,1}}, \quad b_{j,k} = 1  \\
    \end{aligned},
  \right.
\end{equation}
where $b_{j,k} \odot C_j$ is defined as a \textit{switching function}.
If the input bit is $b_{j,k} = 0$, the transformed element is $u_{j,k} = \alpha^{l_{j,0}}$;
otherwise, $u_{j,k}$ is equal to $u_{j,k} = \alpha^{l_{j,1}}$.

Let the input \textit{bit-block} of $J$ users at the $k$-th component denote by ${\bf b}[k]$, 
i.e., ${\bf b}[k] = (b_{1,k}, b_{2,k}, \ldots, b_{J,k})$, where $0 \le k < K$.
We also call the EP codeword $(u_{1,k}, u_{2,k}, \ldots, u_{J,k})$ in $\Psi_{\rm o,B}$ as the output \textit{element-block} of the $k$-th component of $J$ users, i.e., 
${\bf u}[k] = (u_{1,k}, u_{2,k}, \ldots, u_{J,k})$.

We assign the UD-EP $C_j$ in $\Psi_{\rm o,B}$ to the $j$-th user for $1 \le j \le J=m$. 
Then, the multiplexing of the $k$-th components of $m$ users, i.e., $u_{1,k}, u_{2,k}, \ldots, u_{m,k}$, is calculated as 
\begin{equation} \label{e.w_tdma}
\begin{aligned}
w_k &= u_{1,k} + u_{2,k} + \ldots + u_{m,k} \\
    &\overset{(a)}{=} b_{1,k} \alpha^0 + b_{2,k} \alpha^1 + \ldots + b_{m,k} \alpha^{m-1} 
     = {\bf b}[k] \cdot {\bf G}_{\rm M}^{\bf 1}
\end{aligned},
\end{equation}
where (a) is deduced based on the orthogonal UD-EP code $\Psi_{\rm o,B}$ given by (\ref{Psi_oB})
and Eq. (\ref{F_b2q}).
We call ${\bf G}_{\rm M}^{\bf 1}$ the \textit{full-one generator matrix}, given as 

\begin{equation*}
{\bf G}_{\rm M}^{\bf 1} = \left[ 
\begin{matrix}
\alpha^{0}\\
\alpha^{1}\\
\vdots\\
\alpha^{m-1}
\end{matrix}
\right] =
\left[ 
\begin{matrix}
1 & 0 & \ldots & 0\\
0 & 1 & \ldots & 0\\
\vdots & \vdots &\ddots &\vdots\\
0 & 0 & \ldots & 1\\
\end{matrix}
\right],
\end{equation*}
which is an $m \times m$ identity matrix. 
Thus, based on the orthogonal UD-EP code $\Psi_{\rm o,B}$, the proposed system is \textit{a type of time division multiple-access in finite-field (FF-TDMA)}, in which the outputs of the $m$ users completely occupy $m$ locations in an $m$-tuple.

\vspace{-0.1in}
\subsection{Orthogonal Encoding of an Error-Correcting Code}

For $0 \le j < m$, we form the following element-sequence 
${\bf u}_j = (u_{j,0}, u_{j,1}, \ldots, u_{j,k},\ldots, u_{j,K-1})$,                
which is then multiplied $\alpha^j$, given by
\begin{equation} \label{e2.11}
    {\bf u}_j \cdot \alpha^j \triangleq (u_{j,0} \alpha^j, u_{j,1} \alpha^j, \ldots, 
    u_{j,k} \alpha^j, \ldots, u_{j,K-1} \alpha^j).   
\end{equation}

Next, the $m$ element-sequences, i.e., ${\bf u}_0 \cdot \alpha^0, {\bf u}_1 \cdot \alpha^1, \ldots,\\ {\bf u}_{m-1} \cdot \alpha^{m-1}$, can form an FFSP sequence as follows: 
\begin{equation} \label{e2.12}
    {\bf w} =  
    {\bf u}_0 \cdot \alpha^0 + {\bf u}_1 \cdot \alpha^1 + \ldots + 
    {\bf u}_j \cdot \alpha^j + \ldots + {\bf u}_{m-1} \cdot \alpha^{m-1}, 
\end{equation}
where ${\bf w} = (w_0, w_1, \ldots, w_{K-1})$, and $w_k$ is an $m$-tuple.
It indicates that a sequence ${\bf w}$ over GF($2^m$) can be decomposed into $m$ element-sequences ${\bf u}_0, {\bf u}_1, \ldots, {\bf u}_{m-1}$.

Let ${\bf G}_{gc}$ be the generator matrix of a binary $(N, Km)$ linear block code ${\mathcal C}_{gc}$. 
Then, we encode $\bf w$ into a codeword $\bf v$ in ${\mathcal C}_{gc}$ using the generator ${\bf G}_{gc}$, i.e.,
  ${\bf v} = {\bf w} \cdot {\bf G}_{gc} = (v_0, v_1, v_2, \ldots, v_{N-1})$.
It is able to derive that

\vspace{-0.1in}
\begin{small}
\begin{equation*} \label{e2.13}
  \begin{aligned}
{\bf v} =& {\bf w} \cdot {\bf G}_{gc} 
        = ({\bf u}_0 \cdot {\bf G}_{gc}) \alpha^0 \oplus \ldots \oplus 
           ({\bf u}_{m-1} \cdot {\bf G}_{gc}) \alpha^{m-1} \\
        =& {\bf v}_0 \alpha^0 \oplus {\bf v}_1 \alpha^1 \oplus \ldots \oplus {\bf v}_{m-1}\alpha^{m-1}.    
  \end{aligned}
\end{equation*}
\end{small}

\vspace{-0.1in}
where ${\bf v}_j = {\bf u}_j {\bf G}_{gc}$ is the codeword of ${\bf u}_j$ for $0 \le j < m$. The codeword $\bf v$ in orthogonal form is referred to as \textit{orthogonal encoding} of the message $\bf w$, indicating the codeword $\bf v$ is the FFSP of the codewords ${\bf v}_0, {\bf v}_1, \ldots, {\bf v}_{m-1}$.


\section{An FFMA System in a GMAC}
This section presents an FFMA system over a GMAC, based on the orthogonal UD-EP code $\Psi_{\rm o,B}$, i.e.,
  $\Psi_{\rm o,B} = \{C_1, C_2, \ldots, C_j, \ldots, C_J \}$,
where $C_j = \alpha^{j-1} \cdot C_{\rm B}$ for $1 \le j \le J \le m$.
The UD-EP $C_j$ is assigned to the $j$-th user. 

At the transmitter, there are three modules, which are the EP code encoder denoted by ${\rm F}_{{\rm B}2q}$, the channel code encoder ${\bf G}_{gc}$, and the finite-field to complex-field transform function denoted by $\rm F_{F2C}$ which can be realized by modulation.
In fact, the combination of EP code and channel code can be regarded as a \textit{cascade code}, which is used to implement \textit{mulituser channel code}.
At the receiving end, it explores the reverse operations, which are demodulation denoted by $\rm F_{C2F}$, channel code decoder, and EP code decoder denoted by ${\rm F}_{q2{\rm B}}$.

\vspace{-0.1in}
\subsection{Transmitter of a Sparse-form FFMA System}

The transmitter of the $j$-th user for $1 \le j \le J$ maps each bit-sequence ${\bf b}_j$ uniquely into an element-sequence ${\bf u}_j = (u_{j,0},u_{j,1},\ldots, u_{j,k},\ldots, u_{j,K-1})$ by ${\rm F}_{{\rm B}2q}$, i.e., $u_{j,k} = {\rm F}_{{\rm B}2q}(b_{j,k})$, 
which is determined by the EP $C_j = \alpha^{j-1} \cdot C_{\rm B}$.
For $0 \le k < K$, the $k$-th element $u_{j,k}$ of ${\bf u}_j$ is represented by its corresponding $m$-tuple representation over GF($2$),
i.e., $u_{j,k} =(u_{j,k,0}, u_{j,k,1},\ldots, u_{j,k,i},\ldots, u_{j,k,m-1})$, where $0 \le i < m$.
Then, the $m$-tuple form of $u_{j,k}$ is
\setcounter{equation}{13}
\begin{equation} \label{e.u_j_k}
{u}_{j,k} = {\rm F}_{{\rm B}2q}(b_{j,k}) = (0,\ldots, 0, b_{j,k}, 0,\ldots, 0),
\end{equation} 
and the $i$-th component of $u_{j,k}$ is 
\begin{equation} \label{e.u_j}
u_{j,k,i} =
\left\{
  \begin{matrix}
    b_{j,k}, & i = j-1\\
    0,       & i \neq j-1 \\
  \end{matrix}. \right.
\end{equation}
For a large $m$, $u_{j,k}$ is a sparse vector with only one element at the $(j-1)$-th location, and the other locations are all $0$s.
For the binary form of the element-sequence ${\bf u}$, its length is equal to $Km$.

Next, the element-sequence ${\bf u}_j$ of the $j$-th user is encoded into a codeword ${\bf v}_j$ of a binary $(N, mK)$ linear block code ${\mathcal C}_{gc}$.
Suppose the $mK \times N$ generator matrix of ${\mathcal C}_{gc}$ is in systematic form, 
defined by ${\bf G}_{gc,sym}$.
Then, the codeword ${\bf v}_j$ is of the following form
\begin{equation*}
  {\bf v}_j = {\bf u}_j \cdot {\bf G}_{gc,sym}
  = ({\bf u}_j, {\bf v}_{j, {\rm red}}),
\end{equation*}
where ${\bf v}_{j,{\rm red}}$ is the parity (or called redundancy) block, and the subscript ``red'' stands for ``redundancy''.
In the systematic form, the codewords of $m$ users can be arranged in an $m \times N$ codeword matrix
${\bf V} = [{\bf v}_1, {\bf v}_2, \ldots, {\bf v}_j, \ldots, {\bf v}_m]^{\mathrm T}$,
where $1 \le j \le m$ and $J = m$.
The codeword matrix ${\bf V}$ can be divided into two sections, information section ${\bf U}$ and parity section ${\bf E}$, given as

\vspace{-0.15in}
\begin{small}
\begin{equation} \label{e.TM_sparse}
  \begin{aligned}
  {\bf V} =
  \left[
  \begin{matrix}
  {\bf v}_1 \\
  {\bf v}_2 \\
   \vdots   \\
  {\bf v}_{m} 
  \end{matrix}
  \right] 
  =\left[
  \begin{array}{cccc:c}
  u_{1,0} & u_{1,1} & \ldots & u_{1,K-1} & \textcolor{blue}{{\bf v}_{1,\rm red}}\\
  u_{2,0} & u_{2,1} & \ldots & u_{2,K-1} & \textcolor{blue}{{\bf v}_{2,\rm red}}\\
  \vdots  & \vdots  & \ddots &  \vdots   & \vdots  \\
  u_{m,0} & u_{m,1} & \ldots & u_{m,K-1} & \textcolor{blue}{{\bf v}_{m,\rm red}}\\
  \end{array}
  \right]. 
  \end{aligned}
\end{equation}
\end{small}

\vspace{-0.05in}
The information section ${\bf U}$ is a $1 \times K$ array, i.e.,
${\bf U} = [{\bf U}_0, {\bf U}_1, \ldots, {\bf U}_k, \ldots, {\bf U}_{K-1}]$,
and ${\bf U}_k$ for $0 \le k < K$ is an $m \times m$ matrix.
For the information section ${\bf U}$, $J$ information bits $b_{1,k}, b_{2,k}, \ldots, b_{J,k}$ from the $J$ users are lying on the main diagonal of $ {\bf U}_k$.
The parity section ${\bf E}$ is an $m \times (N-mK)$ matrix which consists of all the parity-check bits formed based on the generator matrix, i.e., ${\bf E} = [{\bf v}_{1,\rm red}, {\bf v}_{2,\rm red}, \ldots, {\bf v}_{m,\rm red}]^{\rm T}$.

To support massive users with short packet transmission scenario, $m$ may be very large, indicating ${\bf V}$ is a sparse matrix. In this case, we call ${\bf V}$ a \textit{sparse codeword matrix}, and the corresponding FFMA system is referred to as \textit{sparse-form FFMA (SF-FFMA)}.

Then, each codeword ${\bf v}_j$ is modulated by BPSK signaling and mapped to a complex-field signal sequence ${\bf x}_j \in {\mathbb C}^{1 \times N}$, i.e., 
${\bf x}_j = (x_{j,0}, x_{j,1}, \ldots, x_{j, n}, \ldots, x_{j, N-1})$. 
For $0 \le n < N$, the $n$-th component $x_{j, n}$ is given by
\begin{equation}  \label{e.x_j}
{x}_{j,n} = {\rm F}_{\rm F2C}(v_{j,n}) = 2 {v}_{j,n} - 1,
\end{equation}
where $x_{j,n} \in \{-1, +1\}$. 
The mapping from ${\bf v}_j$ to ${\bf x}_j$ is regarded as 
\textit{finite-field to complex-field transform (F2C)}, 
denoted by $\rm F_{F2C}$. 
Then ${\bf x}_j$ is sent to a GMAC.

\vspace{-0.1in}
\subsection{Receiver of a Sparse-form FFMA System}

At the receiving end, the received signal sequence ${\bf y} \in {\mathbb C}^{1 \times N}$ is the combined outputs of the $J$ users plus noise, i.e.,

\vspace{-0.1in}
\begin{small}
\begin{equation} 
{\bf y} = \sum_{j=1}^{J} {\bf x}_j + {\bf z} = {\bf r} + {\bf z},
\end{equation}
\end{small}

\vspace{-0.1in}
where ${\bf z} \in \mathbb{C}^{1 \times N}$ is an AWGN vector
with ${\mathcal N}(0, N_0/2)$.
The sum in ${\bf y}$ is called \textit{complex-field sum-pattern (CFSP)} signal sequence, i.e., ${\bf r} = (r_0, r_1, \ldots, r_n, \ldots, r_{N-1}) \in \mathbb{C}^{1 \times N}$, 
which is the sum of the $J$ modulated signal sequences ${\bf x}_1, {\bf x}_2,\ldots, {\bf x}_J$,
i.e., $r_{n} = \sum_{j=1}^J x_{j,n}$.

To separate the superposition signals, the classical method is to utilize \textit{multiuser interference cancellation} algorithms.
In our paper, we present another solution based on the USPM structural property of the orthogonal UD-EP code $\Psi_{\rm o,B}$.
We only need to map each CFSP signal $r_n$ to an FFSP symbol $v_n$, i.e., $r_n \mapsto v_n$, and then the bit-sequences of $J$-user can be recovered by decoding the EP codeword.

Thus, the first step of the detection process is to transform the received CFSP signal sequence 
${\bf r} = (r_0, r_1, \ldots, r_n, \ldots, r_{N-1})$ into its corresponding FFSP codeword sequence 
${\bf v} = (v_0, v_1, \ldots, v_n, \ldots, v_{N-1})$ 
by a \textit{complex-field to finite-field (C2F)} transform function $\rm F_{C2F}$, 
i.e., ${v}_n = {\rm F_{C2F}}({r}_n)$ for $0 \le n < N$. 
We find the following facts of the CFSP ${r}_n$ and FFSP ${v}_n$.
\begin{enumerate}
\item
The value of CFSP $r_{n}$ is determined by the number of users who send ``$+1$'' and the number of users who send ``$-1$''. Thus, the set of $r_{n}$'s values in ascendant order is $\Omega_r = \{-J, -J+2, \ldots, J-2, J\}$, in which the difference between two adjacent values is $2$. 
The total number of $\Omega_r$ is equal to $|\Omega_r| = J+1$. 
\item
The value of $v_{n}$ is uniquely determined by the number of $(1)_2$ bits coming from the $J$ users. 
If there are odd number of bits $(1)_2$ from the $J$ users, 
then $v_{n} = (1)_2$; otherwise, $v_{n} = (0)_2$. 
The corresponding FFSP set $\Omega_v$ of $\Omega_r$ is $\Omega_v = \{0, 1, 0, 1, \ldots\}$, 
in which $(0)_2$ and $(1)_2$ appear alternatively. 
\item
Let $C_J^\iota$ denote the number of users which send ``+1''. 
The values of $\iota$ are from $0$ to $J$. 
When $\iota = 0$, it indicates that all the $J$ users send $(0)_2$. 
If $\iota$ increases by one, the number of $(1)_2$ bits coming from $J$ users increases by one accordingly. 
Hence, the probabilities of the elements in $\Omega_r$ are
  ${\mathcal P}_r = \left\{ {C_J^0}/{2^J}, {C_J^{1}}/{2^J},\ldots, {C_J^{J-1}}/{2^J}, {C_J^J}/{2^J} \right\}$.
\end{enumerate}

Based on the relationship between $r_{n}$ and $v_{n}$, we can calculate the posterior probabilities ${\rm Pr}(v_{n} = 0|y_{n})$ and ${\rm Pr}(v_{n} = 1|y_{n})$ for decoding $\bf y$ \cite{FFMA}. 

When the generator matrix ${\bf G}_{gc,sym}$ in systematic form is utilized,
the decoded FFSP codeword sequence $\hat{\bf v}$ can be expressed as $\hat{\bf v} = ({\hat{\bf w}}, {\hat{\bf v}}_{\rm red})$, where ${\hat{\bf w}}$ and ${\hat{\bf v}}_{\rm red}$ are the detected FFSP sequence and parity block, respectively.

After removing the parity block ${\hat{\bf v}}_{\rm red}$ from $\hat{\bf v}$, the detected FFSP sequence ${\hat{\bf w}}$ can be divided into $K$ FFSP blocks, and each block consists of $m$ bits formed an $m$-tuple,
i.e., ${\hat{\bf w}} = ({\hat w}_0, {\hat w}_1, \ldots, {\hat w}_k,\ldots, {\hat w}_{K-1})$, 
where ${\hat w}_k = ({\hat w}_{k,0}, {\hat w}_{k,1}, \ldots, {\hat w}_{k,i}, \ldots, {\hat w}_{k,m-1})$, with $0 \le k < K$ and $0 \le i < m$.
Finally, we separate the detected FFSP block ${\hat{w}}_k$ into $J$ bits as 
$\hat{b}_{1,k}, \hat{b}_{2,k}, \ldots, \hat{b}_{j,k}, \ldots, \hat{b}_{J,k}$ by using an inverse transform function ${\rm F}_{q2{\rm B}}$.
According to (\ref{e.w_tdma}),
we know that the $j$-th user is assigned the EP $C_j = \alpha^{j-1} \cdot C_{\rm B}$.
Thus, to recover the transmit bit information of the $j$-th user, the inverse transform function ${\rm F}_{q2{\rm B}}$ is given as
\begin{equation}
  {\hat b}_{j,k} = {\rm F}_{q2{\rm B}}({\hat w_k}) = {\hat w}_{k,j-1},
\end{equation}
where ${\hat w}_{k,j-1}$ is the $(j-1)$-th component of ${\hat w}_k$. 

After recovering the transmit bit information, we can directly obtain the active users' information from the field-field.
For instance, if ${\hat b}_{j,k} = 0$ for $0 \le k < K$, it can be inferred that the $j$-th user is not active. This phenomenon is appealing for the sourced RA, since it avoids the necessity of addressing the active user detection issue.

\vspace{-0.1in}
\subsection{A Diagonal-form FFMA System}

Recall the \textit{sparse form codeword matrix} ${\bf V}$ given by (\ref{e.TM_sparse}),
for the $1 \times K$ information section ${\bf U}$ of ${\bf V}$,
each entry ${\bf U}_k$ of ${\bf U}$ is an $m \times m$ matrix, and the $m$ information bits $b_{1,k}, b_{2,k}, \ldots, b_{j,k}, \ldots, b_{m,k}$ are located on the main diagonal of $ {\bf U}_k$, 
and the other locations are all $0$s,
where $0 \le k < K$ and $1 \le j \le J \le m$.

Permuting the columns of the $1 \times K$ information array ${\bf U}$ of ${\bf V}$ and rearranging the bit sequences, we can obtain an $m \times m$ information array ${\bf U}_{\rm D}$,
in which ${\bf b}_1, {\bf b}_2, \ldots, {\bf b}_j,\ldots, {\bf b}_{m}$ are lying on the main diagonal of ${\bf U}_{\rm D}$ and the other entries are all $\bf 0$s, given by
\begin{equation}
  {\bf U}_{\rm D} =
  \left[
  \begin{matrix}
  {\bf u}_{1, \rm D} \\
  {\bf u}_{2, \rm D} \\
   \vdots   \\
  {\bf u}_{m, \rm D} 
  \end{matrix}
  \right]
  =
  \left[
  \begin{matrix}
    {\bf b}_1 &           &        &            \\
              & {\bf b}_2 &        &            \\
              &           & \ddots &            \\
              &           &        & {\bf b}_m  \\
 
  \end{matrix}
  \right],
\end{equation}
where ${\bf b}_1, {\bf b}_2, \ldots, {\bf b}_j,\ldots, {\bf b}_{m}$ are $1 \times K$ vectors,
and the subscript ``D'' stands for ``diagonal''.
Let ${\bf u}_{j, \rm D}$ denote as a $1 \times mK$ information sequence of the $j$-th user, 
i.e., ${\bf u}_{j, \rm D} = ({\bf 0}, \ldots, {\bf 0}, {\bf b}_j, {\bf 0}, \ldots, {\bf 0})$,
in which the $1 \times K$ bit-sequence of the $j$-th user ${\bf b}_j$ is located at the $(j-1)$-th entry of ${\bf u}_{j, \rm D}$ and ${\bf 0}$ is a $1 \times K$ zero vector.
Note that, if $J < m$, the $(m-J)$ rightmost entries of ${\bf U}_{\rm D}$ are all zeros, 
forming an $(m-J) \times (m-J)K$ zero matrix.

Then, we encode ${\bf u}_{j, \rm D}$ by the generator matrix ${\bf G}_{gc, sym}$ in systemic form, and obtain the codeword ${\bf v}_{j,\rm D}$ of ${\bf u}_{j, \rm D}$, i.e., 
${\bf v}_{j, \rm D} = ({\bf u}_{j,\rm D}, {{\bf v}_{j,\rm D, red}})$,
where ${\bf v}_{j,\rm D,red}$ is the parity block of ${\bf v}_{j, \rm D}$,
which is a $(N-mK)$-tuple.
The codewords of ${\bf v}_{1,\rm D}, {\bf v}_{2,\rm D}, \ldots, {\bf v}_{J,\rm D}$ together can form a \textit{diagonal form} codeword matrix ${\bf V}_{\rm D}$, given as follows:

\vspace{-0.1in}
\begin{small}
\begin{equation} \label{e_V_D}
  {\bf V}_{\rm D} =
  \left[
  \begin{matrix}
  {\bf v}_{1, {\rm D}} \\
  {\bf v}_{2, {\rm D}} \\
   \vdots   \\
  {\bf v}_{m, {\rm D}} 
  \end{matrix}
  \right]
  =
  \left[
  \begin{array}{cccc:c}
    {\bf b}_1 &           &        &             & \textcolor{blue}{{\bf v}_{1, {\rm D, red}}}\\
              & {\bf b}_2 &        &             & \textcolor{blue}{{\bf v}_{2, {\rm D, red}}}\\
              &           & \ddots &             & \vdots\\
              &           &        & {\bf b}_m   & \textcolor{blue}{{\bf v}_{m, {\rm D, red}}}\\
 
  \end{array}
  \right],
\end{equation}
\end{small}

consisting of an $m \times m$ diagonal array ${\bf U}_{\rm D} = {\rm diag}({\bf b}_1, {\bf b}_2, \ldots, {\bf b}_m)$ and an $m \times 1$ parity block ${\bf E}_{\rm D}$, where ${\bf E}_{\rm D} = [{\bf v}_{1,{\rm D,red}}, {\bf v}_{2,{\rm D,red}}, \ldots, {\bf v}_{m,{\rm D,red}}]^{\rm T}$.

From the codeword ${\bf v}_{j,\rm D}$, it is found that the useful vectors 
(i.e., ${\bf b}_j$ and ${\bf v}_{j,{\rm D,red}}$) 
are only located at the $(j-1)$-th entry of ${\bf u}_{j, {\rm D}}$ and the parity-check entry ${\bf v}_{j,{\rm D, red}}$, and the other entries are all zeros.
Thus, for the diagonal-form codeword ${\bf v}_{j, {\rm D}}$, we can directly modulate and transmit the shorten codeword, defined by ${\bf v}_{j,{\rm D,S}} = ({\bf b}_j, {\bf v}_{j,\rm D, red})$, instead of ${\bf v}_{j,{\rm D}}$, to reduce the transmit power.
The subscript ``S'' indicates ``shorten''.
The codeword length of ${\bf v}_{j,{\rm D, S}}$ is equal to $N - (m-1)K$.

The FFMA system based on the shorten codeword ${\bf v}_{j,{\rm D,S}}$ is referred to as \textit{diagonal-form FFMA (DF-FFMA)} system.
For a short packet transmission, e.g., $K$ is an extremely small number, 
the length of ${\bf v}_{j,{\rm D, S}}$ is approximately equal to the length of parity block ${\bf v}_{j,{\rm D, red}}$.

The permuting and rearranging operations do not affect the properties of FFMA system.
Hence, the DF-FFMA system can be decoded like the aforementioned SF-FFMA system. 
For a DF-FFMA system, the bit-sequences of $J$ users are transmitted in orthogonal mode and the $J$ parity blocks are appended to them.
Obviously, the DF-FFMA system is appealing for the massive users with short data packet transmission scenario, which can obtain a large coding gain from all the massive users with a low-complexity decoder.

\vspace{-0.1in}
\subsection{A Polarization-adjusted FFMA System}

Now, we present a specific configuration of the DF-FFMA, referred to as \textit{polarization-adjusted (or power allocation) DF-FFMA (PA-DF-FFMA)} or simply \textit{PA-FFMA}.

Let $P_{avg}$ denote the average transmit power per symbol. For an $(N, mK)$ channel code $\mathcal{C}_{gc}$, the total transmit power per user is $N P_{avg}$.
In the SF-FFMA system, the entire transmit power of $N P_{avg}$ is utilized. In contrast, the DF-FFMA system only utilizes a portion of this power, specifically $(N - mK + K) \cdot P_{avg}$, leaving $(m-1)K \cdot P_{avg}$ unused.
The proposed PA-FFMA reallocates this unused power. 
By optimizing power allocation, the channel capacity is adjusted according to the assigned power, resulting in a form of \textit{polarization} \cite{PAC_1,PAC_2,PAC_3}. 

The shortened codeword ${\bf v}_{j,{\rm D,S}} = ({\bf b}_j, {\bf v}_{j,\rm D, red})$ of the $j$-th user is consisted of an information block ${\bf b}_{j}$ and a parity block ${\bf v}_{j, \rm D, red}$. 
Assume that the power assigned to the information block per symbol is $\mu_1 P_{avg}$, while the power allocated to the parity block per symbol is $\mu_2 P_{avg}$.
Then, the power allocation (or called polarization-adjusted) conditions are given as

\vspace{-0.1in}
\begin{small}
\begin{equation}
  \begin{aligned}
  C 1&: N P_{avg} = K \cdot (\mu_1 P_{avg}) + (N-mK) \cdot (\mu_2 P_{avg}),\\
  C 2&: 1  \le \left(\mu_{\rm pas} = \frac{\mu_1}{\mu_2}\right) \le m,
  \end{aligned}
\end{equation}
\end{small}

where $\mu_{\rm pas} = \frac{\mu_1}{\mu_2}$ is defined as \textit{polarization-adjusted scaling factor (PAS)}.
Condition $C 1$ ensures that the total transmit power remains constant. 
Condition $C 2$ specifies that the information block receives more power than the parity block. Ideally, we aim to maximize the reliability of the information block by assigning as much power as possible to it.
Additionally, considering the unused power is $(m-1)K \cdot P_{avg}$, the maximum power that can be assigned to the information block per symbol is $m \cdot P_{avg}$, which is equivalent to repeating the information block per symbol $m$ times.
With the increased power allocated to the information block, its reliability improves, leading to a higher degree of polarization. 
Nevertheless, the decoding (or detection) algorithm of PA-FFMA plays a critical role in influencing error performance. For a more detailed discussion, please refer to \cite{FFMA}.

\section{Simulation results}

Now, we show the error performances over a GMAC.
Suppose the frame size (or DoF) is $N = 6000$, each user transmits $K = 10$ bits, and the numbers of total users are $J = 1, 100$ and $300$, respectively. 
We set $\mu_{\rm pas} = 300$ for the PA-FFMA system.

For the SF-FFMA system, each user occupies the entire frame.
In the the DF-FFMA systems, the frame structure consists of $J$ {information blocks} and one parity block. 
For the slotted ALOHA scheme, we assume that the number of time slots is equal to the number of total users, and repetition code is used for error control. Hence, the blocklength of each time slot is $\frac{N}{J}$, and each bit is repeated $\frac{N}{JK}$ times.

\begin{figure}[t] 
  \centering
  \includegraphics[width=0.35\textwidth]{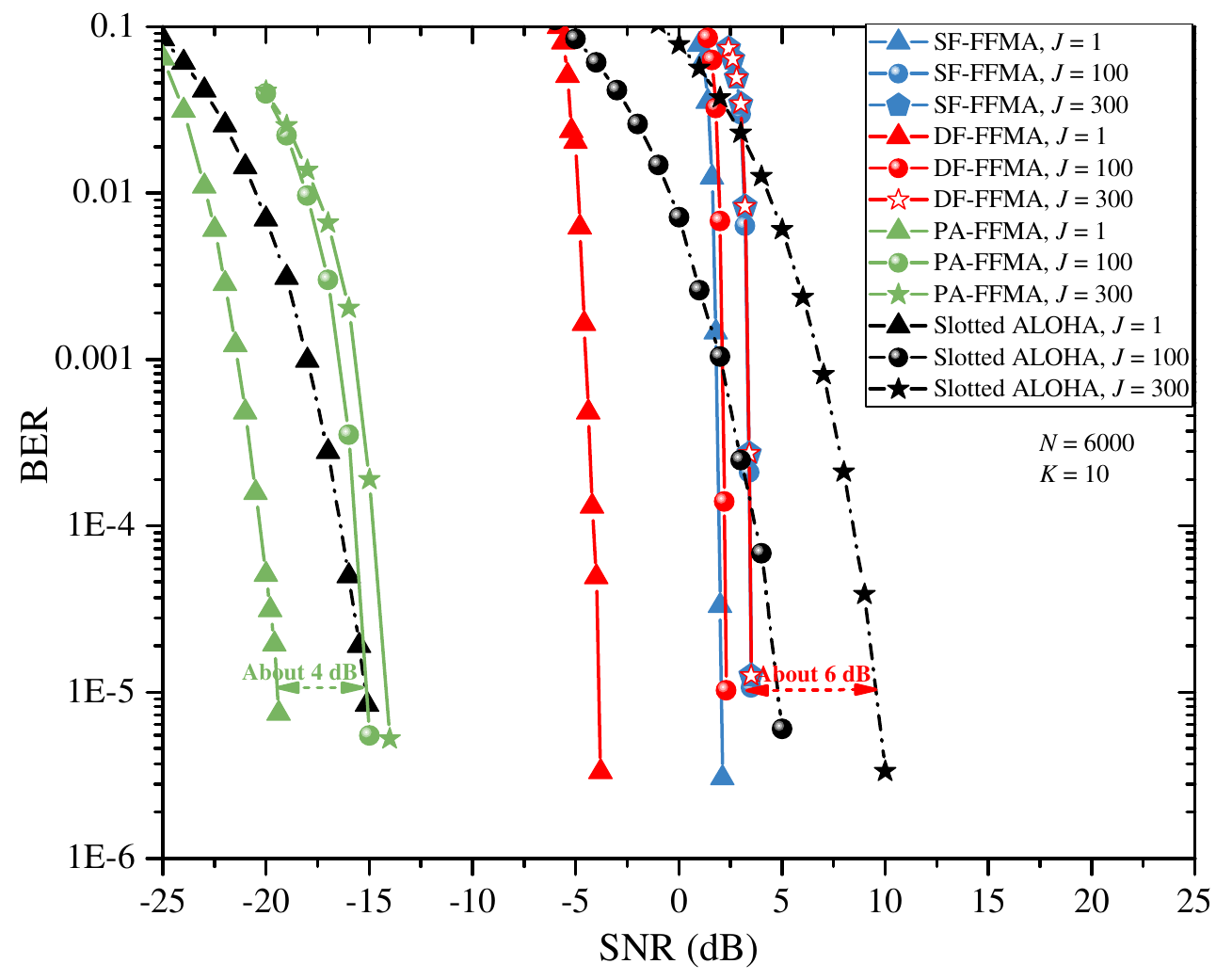}
  \caption{BER performances of different systems over a GMAC. The proposed FFMA systems are used a binary $(6000, 3000)$ LDPC code ${\mathcal C}_{gc}$ for error control, and the slotted ALOHA utilizes repetition code for error control, where $N = 6000$, $K = 10$ bits, $\mu_{\rm pas} = 300$, and $J = 1, 100, 300$.}
  \vspace{-0.2in}
\end{figure}

From Fig. 1, we observe that the BER decreases as the number of users increases. For the SF-FFMA systems, the BERs for cases where $J = 100$ and $J = 300$ are nearly identical. When the BER is $P_b = 10^{-5}$, the difference between the BERs of $J = 1$ and $J = 300$ is only $1.5$ dB.
We then compare the BER performance between the SF-FFMA and DF-FFMA systems. 
Under the same simulation conditions, the DF-FFMA system achieves a lower BER than the SF-FFMA, as the default bits, which are $0$s, are available at the receiving end. However, as the number of users increases, the difference in BER between the SF-FFMA and DF-FFMA systems diminishes. For $J = 300$, both systems exhibit the same BER.
Additionally, it is found that the PA-FFMA system provides significantly better BER performance than both the SF-FFMA and DF-FFMA systems. This improvement is attributed to a polarization gain of approximately $10 \log_{10}(\mu_{\rm pas}) \approx 24.77$ dB.

Next, we compare our proposed SF-FFMA and DF-FFMA systems with the slotted ALOHA system. 
When $J = 1$, slotted ALOHA with repetition coding outperforms both the SF-FFMA and DF-FFMA systems in terms of BER due to a coding gain of approximately $10 \log_{10}(\frac{N}{JK}) \approx 27.78$ dB. Besides, the proposed PA-FFMA system offers an additional coding gain of approximately $4$ dB over the slotted ALOHA system, as it combines both polarization gain from power and coding gain from the entire DoF.

For a large number of users, i.e., $J \ge 100$, all three configurations of the proposed FFMA systems can offer much better BER performance than the slotted ALOHA system. Specifically, at $P_b = 10^{-5}$ and $J = 300$, the SF-FFMA (or DF-FFMA) systems provide a coding gain of approximately 6 dB over the slotted ALOHA system, further validating the ability of the proposed FFMA systems to enhance BER performance in scenarios with massive user counts.

\section{Conclusions}
In this paper, we propose a novel FFMA technique designed to support a massive number of users with short packet traffic, addressing the challenges of FBL in multiuser reliable transmission. Unlike traditional complex-field MA systems, we utilize EPs as virtual resources to distinguish between users. We introduce a type of orthogonal UD-EP code, denoted as $\Psi_{\rm o, B}$, constructed over GF($2^m$), which can form TD-FFMA mode. Simulation results demonstrate that the proposed FFMA systems can support a large number of users while maintaining favorable error performance in a GMAC.

\newpage

\vfill

\begin{thebibliography}{99}








\bibitem{6G}
C. -X. Wang et al., ``On the Road to 6G: Visions, Requirements, Key Technologies, and Testbeds,'' \textit{IEEE Communications Surveys \& Tutorials}, vol. 25, no. 2, pp. 905-974, Secondquarter 2023.



\bibitem{UMA_PZ}
P. Fan et al., ``Random access for massive Internet of things: current status, challenges and opportunities,'' \textit{Journal on Communications}, vol. 42, no. 4, pp. 1-21, April 2021.


\bibitem{UMA_2022}
Y. Li et al., ``Unsourced multiple access for 6G massive machine type communications,'' \textit{China Communications}, vol. 19, no. 3, pp. 70-87, March 2022.



\bibitem{MIT_2017}
Y. Polyanskiy, ``A perspective on massive random-access,'' \textit{IEEE International Symposium on Information Theory-Proceedings}, 2017, pp. 2523–2527. 



\bibitem{SourcedRA_1}
M. Ke, Z. Gao, M. Zhou, D. Zheng, D. W. K. Ng and H. V. Poor, ``Next-Generation URLLC With Massive Devices: A Unified Semi-Blind Detection Framework for Sourced and Unsourced Random Access,'' \textit{IEEE Journal on Selected Areas in Communications}, vol. 41, no. 7, pp. 2223-2244, July 2023.


\bibitem{Yu_UDAS}
Qi-yue Yu, and Ke-xun Song, ``Uniquely Decodable Multi-Amplitude Sequence for Grant-Free Multiple-Access Adder Channels,'' \textit{IEEE Transactions on Wireless communications}, vol. 22, no. 12, pp. 8999-9012, Dec. 2023.


\bibitem{SourcedRA_2}
X. Shao, X. Chen, D. W. K. Ng, C. Zhong and Z. Zhang, ``Cooperative Activity Detection: Sourced and Unsourced Massive Random Access Paradigms,'' \textit{IEEE Transactions on Signal Processing}, vol. 68, pp. 6578-6593, 2020.






\bibitem{ALOHA_1}
F. Schoute, ``Dynamic Frame Length ALOHA,'' \textit{IEEE Transactions on Communications}, vol. 31, no. 4, pp. 565-568, April 1983.



\bibitem{Capacity_GMC_2017}
X. Chen, T.-Y. Chen and D. Guo, ``Capacity of Gaussian many-access channels,'' \textit{IEEE Trans. Inf. Theory}, vol. 63, no. 6, pp. 3516-3539, Jun. 2017.




\bibitem{Capacity_GMC_2021}
R. C. Yavas, V. Kostina and M. Effros, ``Gaussian Multiple and Random Access Channels: Finite-Block length Analysis,'' \textit{IEEE Transactions on Information Theory}, vol. 67, no. 11, pp. 6983-7009, Nov. 2021.




\bibitem{PAC_1}
E. Arıkan, ``On the Origin of Polar Coding,'' \textit{IEEE Journal on Selected Areas in Communications}, vol. 34, no. 2, pp. 209-223, Feb. 2016.

\bibitem{PAC_2}
E. Arıkan, ``From sequential decoding to channel polarization and back again,'' 
https://arxiv.org/abs/1908.09594.

\bibitem{PAC_3}
M. Rowshan, A. Burg and E. Viterbo, ``Polarization-Adjusted Convolutional (PAC) Codes: Sequential Decoding vs List Decoding,'' \textit{IEEE Transactions on Vehicular Technology}, vol. 70, no. 2, pp. 1434-1447, Feb. 2021.
























































































\bibitem{LinBook3}
J. Li, S. Lin, K. Abdel-Ghaffar, W. E. Ryan, and D. J. Costello, ``LDPC Code Designs, Constructions, and Unification," Cambridge University Press, 2017.

\bibitem{FFMA}
Qi-yue Yu, Jiang-xuan Li, and Shu Lin,``Finite Field Multiple Access,'' https://arxiv.org/abs/2303.14086v2.





\end{thebibliography}
\end{document}